# High-Coherence and High-frequency Quantum Computing: Going Towards a High-Frequency, High-Coherence and Scalable Quantum Computing Architecture.

Masroor H. S. Bukhari [1,2,3]
[1] Sir Syed Science Park, Saadi Avenue, Clifton, Karachi 75500, Pakistan.
[2] Department of Physics, Faculty of Science, Jazan University,
Jazan 82817-2820, Saudi Arabia.
[3] Physics Division, U. S. Jefferson National Laboratory
and Accelerator Facility, 12000 Jefferson Avenue,
Newport News, VA23606, United States of America.
Email: mbukhari@jazanu.edu.sa, mbukhari@jlab.org.
(ORCID: 0000-0003-3604-3152. WOS Researcher ID: F-8375-2013.)

**Abstract**

High-coherence, fault-tolerant and scalable quantum computing architectures with unprecedented long coherence times, faster gates, low losses and low bit-flip errors may be one of the most viable ways forward to achieve the true quantum advantage. In this context, high-frequency high-coherence (HCQC) qubits with new materials and advanced, high-performance topologies could be a significant step towards making efficient and high-fidelity quantum computing possible, mainly by facilitating higher anharmonicities and thermal suppression, compact size, higher scalability and higher than conventional operating temperatures. Although transmon type qubits are designed and manufactured routinely in the range of a few Giga-Hertz, normally from 4 to 7 GHz (and, at times, up to around 10GHz, with one group reporting an exceptional qubit operation up to 72GHz), achieving higher-frequency operation has challenges and entails special design and manufacturing considerations. This report provides an overview of the foundations of transmon-based quantum computing architectures and presents the proposal and preliminary (limited) design of an 8-qubit transmon (with possible upgrade to up to 72 qubits on a chip) architecture working beyond an operation frequency of 10GHz, as well as presents a new connection topology. The current high-frequency design spans a range of around 11 to 12GHz (with a possible full range of up to 13.5GHz possible at the moment), with a central optimal operating frequency of 11.3 GHz, with the aim to possibly achieve a stable, compact and low-charge-noise operation, as lowest as possible, as per the existing fabrication techniques. The aim is to achieve average relaxation times of up to 1.9ms, in the beginning, with average quality factors of up to $2.75\times10^7$ after trials, while exploiting the new advances in superconducting junction manufacturing, using tantalum and niobium/aluminum/aluminum oxide tri-layer structures on high-resistivity silicon substrates (as reported in recent literature). As an initial transmon-based 11.3/12GHz prototype, being developed at present with a small number of eight qubits on a chip, using a coplanar four-body coupler (Quad-Transmon-Coupler, QTC) architecture, in future, it could be upgraded to accommodate up to 72 qubits on a single chip, with Octa-Transmon-Coupler (OTC) configuration, and with that, one could achieve a possible higher frequency operation up to 20's to 30's of GHz and thermal stability up to 150-200mK in the higher operating temperature regime (which in the current design is limited to 65mK), using a bit advanced chip manufacturing processes and more involved design considerations.

## 1. Introduction

There seems to be a frantic race and an on-going scientific and engineering challenge the world over to build more and more robust and fault-tolerant, high-fidelity quantum bit (qubit) architectures. The core requisites appear to be high qubit density, long-duration quantum states, low error rates, scalability and better error correction algorithms. The transmon qubit architecture [1, 2, 3], which is a type of superconducting charge quantum device evolved from a primitive Cooper box device, offers fast gate implementation, good coherence times, superior scalability, easy coupling and reduced charge noise sensitivity, however, at the same time, it entails some major challenges like the need for extremely low operating temperatures, noise isolation from the environment and technical difficulties to fabricate and implement at higher frequencies (more than tens of GHz). Nevertheless, transmon continues to be the best de facto architecture for simple, scalable and efficient quantum computing at the moment, especially in view of its simple fabrication and high scalability for large-scale computing.

In order to achieve the anticipated quantum advantage from quantum computing, some of the main prerequisites are keeping the phase information preserved and achieving sufficiently long coherence times. It may be possible, in this context, that high-coherence, high-frequency and fast gate computing may be the only way to implement quantum computing units that outperform all classical computing seen in past. Higher frequency qubits (involving higher than the usual <10GHz qubit resonator and readout system frequency) offer many significant benefits over the low-frequency ones, which include but are not limited to, faster gate operations, higher thermal suppression (and possibly higher anharmonicities), higher operating temperatures, scalability (owing to compact qubit-resonator size and occupation area on the chip wafer) and lesser initialization errors, as recently demonstrated in a few studies involving high-frequency architectures [4, 5]. In short, if correctly designed and carefully implemented, the paradigm of High-Coherence Quantum Computing (HCQC) involving high-qubit frequency operation may become the future of quantum computing.

We present here a brief review of the basic elements of a transmon qubit architecture in addition to the proposal of a new design of a transmon array architecture and some relevant guidelines for its operation in the high-frequency regime, in addition to offering a possible enhanced operation, if implemented correctly. The endeavor is a result of long considerations in reaching a viable design by the author as well as taking the best bits of the qubit designs presented elsewhere in the literature, which were pertinent to a conducive high-coherence operation. In terms of the qubit transmon design, we employ new and enhanced qubit and coupler designs and fabrication techniques introduced during the last five years [4, 5, 6, 7, 8], along with some of our own new improvisations, in order to reach a robust high-frequency and high-coherence transmon architecture. The detection and readout section is mainly based upon our previous work on the detection of ultra-weak microwave photons within a microwave cavity with the help of relevant Josephson junction-based devices (for the application of hypothetical dark matter particle searches, as reported in [9] and in our earlier work referred therein), however, in the current design, a new detection and readout technique is employed, using a quantum-limited array of Traveling Wave Parametric Amplification (TWPA) units [10].

The architecture is based upon two distinct improvements, first, it involves the qubits made with Tantalum films using a dry etching process, based upon the original work by two teams [11, 12], and at second, we propose a new transmon-coupler geometry as well as some design considerations (in both the qubit device geometry as well as in the readout amplification chain and methods) for a successful and sustained operation of qubits beyond the 10GHz barrier, initially proposing an 8-qubit architecture with operating frequencies around 11 to 12 GHz (and possibly up to 13.5GHz.) Recent studies have demonstrated that qubits can indeed be manufactured and operated at frequencies up to 72GHz (transcending through the millimeter-wave arena) and can operate in temperatures up to 200mK [5].

Increasing the resonance frequency and thus achieving a higher frequency operation translates to increased noise immunity and qubit anharmonicity, faster gate operation, and reduced thermal activation [5]. In addition, a higher resonance frequency means a more compact design and intelligent packaging, in addition to minimizing dielectric and radiation losses, and thus improved and higher scalability, if properly designed and implemented. In terms of cooling, the high-frequency regime has associated with it simpler system design and the possibility of increasing the operating temperature (whether by using the powerful helium adsorption method with Helium4 ($^{4}$He) as cooling agent (cryogen) or by the usage of the usual dilution refrigeration).

The aim of the devised architecture is thus to achieve high-frequency, long-coherence and fault-tolerant operation (with possibly a small footprint device) to achieve improved coherence and higher scalability as well as improved integration density. Although it is a preliminary effort, but after long evaluations it could be materialized into a working quantum computing solution for solving a myriad number of important computational problems.

## 2. Theory

A transmon, an abbreviation of "Transmission line shunted plasma oscillation qubit" (often abbreviated as "Xmon") is a superconducting tunneling junction based device involving two superconductors separated by a thin barrier (forming a Josephson Junction, JJ), and acts as a high-quality, non-linear microwave oscillator (an emulation of an inductor-capacitor LC-like circuit) [1, 2], serving as a quantum bit (qubit) device, or as the core of a quantum computing experiment, or a Quantum Processing Unit (QPU). It is in reality a non-linear inductance-capacitance (LC) oscillator, based upon the initial idea of a Cooper box [2], where an L-C network is acting as an artificial atomic system with well-defined, quantized energy states (similar to an atom) with a central resonant frequency, which depends upon the chosen L and C values of the circuit (dependent upon the dimensions and structure of the device.) By the virtue of the non-linearity inherent in the JJ, a transmon can create an energy spectrum with uneven energy gaps, unlike a Simple Harmonic Oscillator which has an evenly-spaced energy spectrum. A large shunting capacitor in parallel with the JJ significantly reduces the device's sensitivity to charge noise, providing a quantum leap over the original Cooper box design. A multiple qubit transmon architecture is essentially a system of coupled nonlinear resonator circuits involving a set of a series of Josephson junctions. A transmon differs from a conventional cooper box in the sense that it has a large bias or shunting capacitor ($C_B$) between the two superconductors with its own contributions to the total capacitance of the circuit ($C_\Sigma$). An integral and important accessory to a transmon is a readout resonator (RR), which is typically a coplanar waveguide, capacitively coupled to the former in a dispersive readout scheme. The state of a transmon is readout and a measurement is done with the help of reading out the phase information of microwave pulses (generated from an external RF source) scattered off the transmon in its kept state. A simple transmon architecture is depicted in the Figure 1, where a set of two JJ's and a flux biasing circuit are shown. A configuration in which a set of two parallel Josephson junctions are employed is the usual Superconducting QUantum Interference Device (SQUID), where one can achieve frequency tuning by means of magnetic flux tuning of the energy stored in the junctions (the "flux-tunable qubits"). It can also help one achieve two-qubit gate circuits.

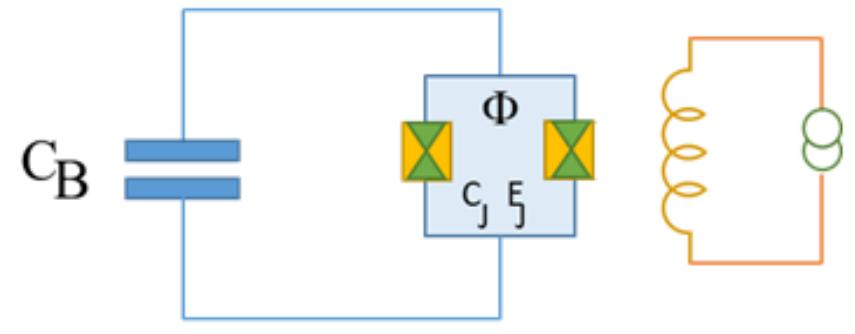

FIGURE 1. A Simple Tunable Transmon Qubit with Two Josephson Junctions and a Large Shunt/Bias Capacitor (and Flux Biasing).

A detailed theoretical treatment of a transmon qubit architecture within the Schrödinger equation-based Hamiltonian formalism of a simple harmonic oscillator and its interaction with external electromagnetic fields, as per the highly successful Jaynes-Cummings model (including that of a qudit, a multiple-transition or multiple-state ($n > 2$) version of a qubit), may be found elsewhere, mainly in the detailed papers by [13, 1, 2, 4], however, it is imperative that pertinent theoretical background of the relevant quantities may be provided here.

Beginning with the description of the energy stored within the LC network of a transmon, one can write an LC Hamiltonian (total energy function) as:

$$H_{LC} = \frac{Q^2}{2C} + \frac{\Phi^2}{2L}. \tag{1}$$

Where $Q$ is the charge on the qubit and $\Phi$ is the magnetic flux.

One can obtain a free Hamiltonian for the LC resonator within a simple harmonic oscillator formalism, by rewriting the Hamiltonian in terms of the quantum raising and lowering ladder operators $(\hat{a}, \hat{a}^+)$ and the qubit's central resonance frequency, $\omega_q$ (not to be confused with the readout resonator frequency, $\omega_r$), and introducing and taking into account the important and central parameter of the Charging energy ($E_C$), as [20]:

$$\widehat{H}_0 = \hbar\omega_q \hat{a}^+ \hat{a} - \frac{E_C}{12}(\hat{a} + \hat{a}^+)^4. \tag{2}$$

Where the qubit's non-linearity parameter ($\kappa$) coupling to the Charging energy, that may be modeled in a number of ways and bears a small value, has been assigned a value of $1/12$ here [13].

The intrinsic resonance frequency, the working frequency, of the transmon resonator (traditionally known as the "Plasma frequency" in physics models) is expressed as:

$$\omega_q = \omega_p = \frac{\sqrt{8E_J E_C}}{\hbar}. \tag{3}$$

Or we can also write it as:

$$\nu_q = \nu_p = \frac{\sqrt{8E_J E_C}}{h}, \tag{4}$$

The transition frequency from one state to the other, for instance from the level "0" to "1", is expressed as:

$$\nu_{01} = \frac{\sqrt{8E_J E_C}}{h} - \frac{E_C}{h}, \tag{5}$$

The Josephson energy, which is a function of the potential energy associated with the junction, and the Charging energy are expressed, respectively, as:

$$E_J = \frac{\Phi_0 I_c}{2\pi}, \tag{6a}$$

$$E_C = \frac{e^2}{2C_\Sigma}, \tag{6b}$$

Where $e$ is the elementary charge (the charge on an electron), $\Phi_0$ is the magnetic flux quantum (another elementary constant, $hc/2e$), and $I_C$ is the critical current across the junction. Here $C_\Sigma$ denotes the collective capacitance, which is in turn, given by:

$$C_\Sigma = C_J + C_B + C_g. \tag{7}$$

Where $C_g$ is the gate capacitance and $C_B$ is the value of the Bias capacitance.

The junction inductance ($L_J$) is expressed as a function of the critical current and elementary charge/magnetic flux as:

$$L_J = \frac{\hbar}{2eI_C} = \frac{\Phi_0}{2\pi I_c}, \tag{8}$$

Following a rotating wave approximation as usual [1, 2], the transformed (rotated) Hamiltonian in the rotated frame of the transmon is written as:

$$\widetilde{H} = -E_J \left[ cos\hat{\widetilde{\Phi}} + \left(1 - \frac{1}{2}\hat{\widetilde{\Phi}}^2\right)\right]. \tag{9}$$

Finally, after solving, an effective Hamiltonian for the transmon in the quantum picture is obtained [20, 14, 15], and one writes it as:

$$\widehat{H}_{Xmon} = 4E_C\left(\hat{n} - n_g\right)^2 - E_J cos\hat{\varphi}, \tag{10}$$

Where $\hat{n}$ stands for the number of cooper pairs transferred across the junction, $\hat{\varphi}$ the (gauge-invariant) phase difference operator (corresponding to the phase difference, between the two superconductors in the junction, that causes the transfer), and $n_g$ denotes the charge offset number operator or operator corresponding to the effective offset charge of the qubit induced by the source, and is conjugate to the $\hat{\varphi}$ operator, expressed as [14]:

$$n_g = \frac{Q_r}{2e} + \frac{C_g V_g}{2e}. \tag{11}$$

With $Q_r$ standing for the offset charge induced by the environment, and $C_g$ and $V_g$ being the gate capacitance and voltage, respectively.

This was the theoretical expression of a stand-alone transmon's energy as given by the Hamiltonian above.

Once a driving electromagnetic field is introduced in the transmon system, for instance, with the help of a train of driving pulses, the total system energy (Hamiltonian) in quantum picture is the sum of the transmon and the time-dependent drive field energy, and is in a time-evolving such as:

$$\widehat{H}_{system}(t) = \widehat{H}_{xmon} + \widehat{H}_{drive}(t), \tag{12}$$

In that case, a coupled electromagnetic field-transmon system Hamiltonian is written in terms of the Josephson junction energy ($E_J$), electromagnetic field energy ($\hbar\omega_p$), driving frequency ($\omega_d$) and the Rabi frequency ($\Omega_{Rabi}$) as [15]:

$$\widehat{H}(t) = \hbar\omega_q \hat{a}^{+}\hat{a} - E_J\left[cos\widehat{\Phi} + \left(1 - \frac{1}{2}\widehat{\Phi}^2\right)\right] + \hbar\Omega_{Rabi}(cos\omega_d t)(\hat{a} + \hat{a}^{+}). \quad (13)$$

Where $\widehat{\Phi}$ is the flux operator, $\omega_q$ is the qubit transition frequency, and $\omega_d$ is the driving microwave pulse frequency.

We now move into the rotating frame of the driving microwave field, similar to the treatment we did for the transmon, and after solving, we find the Hamiltonian for the transmon-field system in the rotated frame of the transmon and transformed frame of the field, respectively, as [15]:

$$\widetilde{H} = -E_J\, cos\widehat{\Phi}_q - E_J\left(1 - \frac{1}{2}\widehat{\Phi}^2{}_q\right). \quad (14)$$

Thus, we obtain an expression for the energy of a transmon-microwave field system for our experiment/computational unit.

In a multi-qubit architecture there are many viable configurations possible whereby qubit transmons can be coupled to common resonators via suitable couplers, with each configuration offering its own benefits and respective shortcomings. In recent times there have been viable working qubit configurations reported in literature with two or three transmons coupled to a single resonator [16, 17, 18, 54].

One can write the Hamiltonian of a multiple transmon system (that evolves in time), for n number of transmons (labelled by index *i*), each one depicted by $\widehat{H}_{T_i}$, along with that of a single resonator, $\widehat{H}_R$, and of their interaction, $\widehat{H_I}$, as follows:

$$\widehat{H}(t) = \sum_{i=1}^{n} \widehat{H}_{T_i}(t) + \widehat{H}_R + \widehat{H_I}, \quad (15)$$

Whereas the Hamiltonian for each individual transmon is written out in terms of the relevant parameters of the charging energy and Josephson energy of each one, and as a function of the superconducting phase (that is conjugate to the energy of the corresponding *ith* transmon) as:

$$\widehat{H}_{T_i}(t) = 4E_{C_i}\{\hat{n}_i - n_{g_i}(t)\}^2 - E_{J_i}cos\hat{\varphi}_i. \quad (16)$$

The Hamiltonian can alternatively be written with the individual energy components of a transmon, viz. the Josephson junction, inductance and capacitance, laid out, and in a simplified manner for a single transmon unit, as summed up by Glazman and Catelani as [];

$$\widehat{H} = 4E_C\left(\frac{1}{I}\frac{d}{d\varphi} - n_g\right)^2 - E_J cos\hat{\varphi} + \frac{1}{2}E_L\left(\hat{\varphi} - 2\pi\frac{\Phi_e}{\Phi_0}\right)^2 \quad (17)$$

This Hamiltonian, as derived by Glazman and Catelani, depicts a large number of superconducting transmon-based qubit types, and, thus, we obtain a general mathematical framework to understand the workings of these devices.

However, it is important to understand that these are ideal expressions and they may and do deviate in actual implementations, especially at high frequencies, where the last resort is engineering simulations. Nevertheless, it provides a pertinent and basic mathematical framework for modeling basic transmon systems.

The figure 2 illustrates a basic transmon and resonator system along with the associated flux, feed and drive lines, acting as quantum buses similar in concept to (but unlike) classical digital computer architecture buses.

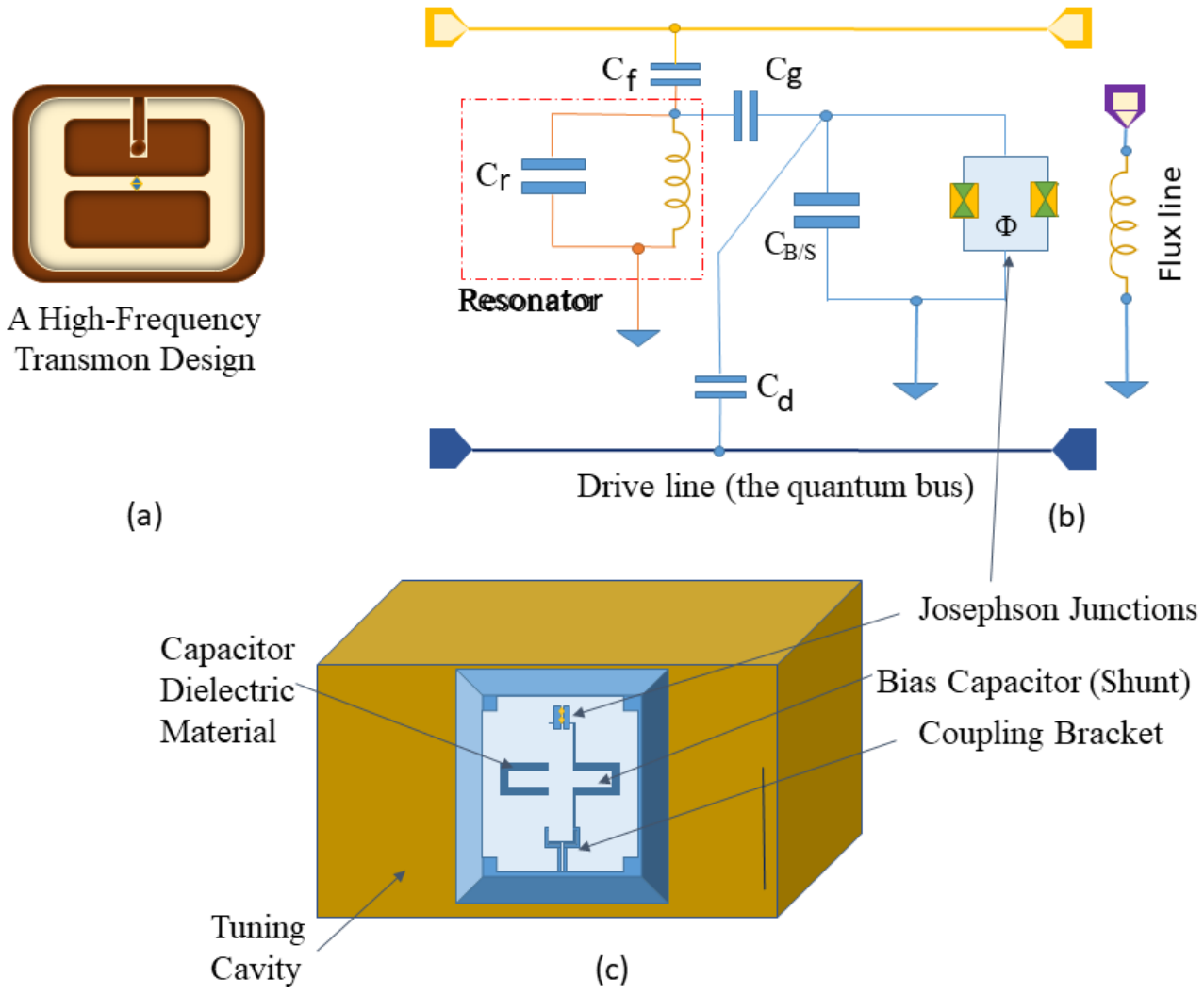

FIGURE 2. A Complete Transmon and Resonator System with various Quantum Buses. (a)A Single Transmon. (b) A Complete Transmon and Resonator Qubit System with Flux Biasing. The Yellow line on the top depicts the Feedline, whereas the Blue/ Black line on the bottom is the Drive line. (c) A Transmon Qubit Enclosed in a Copper Tuning Cavity.

More details and an in-depth theoretical treatment of the involved subject, including various topologies and design variations, such as fluxonium and higher-dimensional flux qubits, as well as details of the underlying SQUID framework, etc., are provided in two excellent treatises on the subject [19, 20, 21].

## 3. Architecture

The High Coherence Quantum Computing (HCQC) may be considered as the quantum computing's ideal paradigm where the qubits retain their coherence and remain stable and coherent for sufficiently long enough times to carry out a large number of complex computing operations before the information stored in an entangled form within the qubits is lost to the environment. While contemplating the factors affecting the coherence, it is the parasitic Two-Level System (TLS) losses that are the primary enemy. Thus, the primary focus of a successful high-coherence high-frequency qubit design and implementation must be on mitigating all possible dielectric and TLS losses in order to preserve coherence.

### *3.1. Anharmonicity engineering*

As one contemplates the background of the losses in a qubit structure, one finds that the design of a high-coherence transmon-based quantum computing architecture hinges upon a few main factors, which primarily include the intrinsic times that a qubit is associated with in going out of coherence, and mainly depend upon a qubit's design topology, intrinsic electrical parameters, thermal performance, noise spectral density and its immunity to the external electromagnetic and incident energetic particle fluxes (in addition to fabrication techniques). These times are, in general, the energy relaxation (or longitudinal relaxation) and overall dephasing times (the $T_1$ and $T_2$ times, respectively) and echo dephasing ($T_2^{echo}$) time, as well as dependent on the gate speed ($T_g$). In particular, the loss of coherence greatly depends upon the $T_1$ and $T_2$ times, and thus, these are the two main benchmarks for qubit performance.

The following relationship between the three times holds;

$$T_2^{echo} \leq T_2 \leq 2T_1. \quad (18)$$

Once a measurement is made, the ensuing echo sequence always helps improve the obtained coherence time by refocusing the slow-occurring noise.

Most importantly, it is the ratio of the coherence times to the gate speed that is paramount in deciding the coherence of a qubit. A $T_2, T_g$ combination (ratio) of around $\sim 10^3 - 10^4$ is desirable in terms of design specifications. For fault-tolerant computing involving gate fidelities of greater than 99.9% one requires more than $\sim 10^4$ gate operations within a particular coherence time. The typical $T_2$ achieved with the high-performance design we contemplate here is $> 30\mu s$ (possibly up to $100\mu s$) whereas the $T_g$ of around $10 - 20ns$ is desirable, so as to achieve the desired rate, however with the help of a careful design we wish to improve upon these times and acquire the optimal rate as required for fault-tolerant computing. Thus, the aim of this design is to maximize as much as possible the coherence by achieving the best possible combination of these, as well as, achieve the longest possible energy relaxation $T_1$ and echo dephasing $T_2^{echo}$ times by means of design and fabrication advancements. Necessary energy relaxation and Hahn Echo experiments [4] shall be carried out to benchmark the $T_1$ and $T_2^{echo}$ times, respectively.

The important and central parameter of qubit anharmonicity ($\alpha$), which is, in fact, nothing but the difference between the two transition frequencies of a qubit, is expressed as:

$$\alpha = \frac{E_C}{h} \approx v_{01} - v_{12}, \quad (19a)$$

In addition, the anharomonicity also depends upon the ratio of the Joseph-son energy to the charging energy.

$$\alpha/2\pi \approx E_J/E_C. \quad (19b)$$

The key to the successful transmon operation, especially in achieving the quantum advantage, is that by maximizing the $E_J/E_C$ ratio, the device becomes virtually immune to the charge noise.

In order to achieve a high-frequency operation, the similar ideology is followed, and an acceptable and optimal $E_J/E_C$ ratio is entailed ($E_J \gg E_C$), which is both not too high, so as to preserve the desired anharmonicity, as well as not too low. Therefore, the aim of an optimal transmon operation is thus to maintain a very high $E_J/E_C$ ratio without which a high-coherence could never be achieved. Moreover, as aptly pointed out by Koch *et al.* [2], one of the great benefits of the transmon architecture qubit, in contrast to the primitive Cooper box configuration, is that (by means of careful design) in the former (transmon qubit) one can strongly suppress charge dispersion to the extent that it is virtually immune to charge noise, and thus, one can greatly reduce the qubit's charge noise sensitivity while at the same time, not sacrificing anharmonicity [2].

The charging energy depends upon the value of the capacitance of the body of the transmon as well as on the parameters of the Josephson junction (JJ), whereas the Josephson energy depends upon the critical current through the JJ.

The values of the $E_J/E_C$ ratio, the shunt capacitance and critical current laid out in the design, as of now, are 100, 47fF, and 240nA, respectively, however, it is difficult to determine them with utmost accuracy. We aim to achieve Josephson energy vs. charging energy ($E_J/E_C$) ratio near 100 in order to reduce charge noise as much as possible and achieve a balance between optimal anharmonicity and noise immunity. We also aim to achieve a bit higher gate fidelity than that achieved in earlier designs with the help of an improved design.

In terms of calculating the Anharmonicity of our qubits, we follow a similar approach by Moretti *et al.* []. The qubit anharmonicity, depicted as $\chi_\alpha$, and its related quantity, the total dispersion (i.e. the dispersive shift as a result of qubit transitions), $\vartheta$, in a simplest form, as a function of the Josephson energy and the two modes with the respective participation energy ratios of the qubit element for two modes $\alpha, \beta$, are given as:

$$\chi_\alpha = \frac{\hbar\omega_\alpha\omega_\beta}{4E_j} p_\alpha p_\beta, \;\; \vartheta_{\alpha\beta} = 2\chi_\alpha. \quad (20)$$

An important variable of central importance in a microwave resonator is the Quality Factor of the resonator, which determines the losses in the resonator owing to energy dissipation, and is generally known as a "Q" value. A high-quality resonator has an extremely high quality value, but it cannot exceed a certain level owing to material limitations.

The longitudinal relaxation rate ($\Gamma_1$), also known as the Energy exchange rate, is inversely related to the time $T_1$, and can be quantified in terms of the quality factor (Q) and qubit transition frequency ($v$) as [30]:

$$\Gamma_1 = \frac{1}{T_1} = \frac{2\pi v}{Q}. \quad (21)$$

The architecture is based upon two distinct improvements, first, it involves the qubits made with Tantalum films, based upon the original work by Place *et al.* [8], and the modifications made by Wang *et al.* [12], especially the usage of a dry etching process in fabrication, and at second, we propose some design considerations (in both the qubit device as well as in the readout chain) for a successful and sustained operation of qubits beyond the 10GHz barrier, initially proposing an 8-qubit architecture with operating frequencies around 11 to 12 GHz (and possibly up to 13.5GHz using the existing design.)

### 3.2. High-frequency transmon design

The two main factors that decide the operating frequency (the natural or resonance frequency) of a transmon qubit, similar to all the LC circuits are the Josephson Inductance ($L_J$) and the Shunt/Bias Capacitance ($C_B$), as illustrated in the Figure 1, the circuit schematic of the basic transmon in our implementation.

The strategy adopted here to achieve a high-frequency operation is one, to achieve a higher junction critical current density ($J_C$), by means of adopting small as possible junction finger dimensions (with compact interdigitized capacitors), a factor on which the frequency largely depends upon, similar to the work carried out by [4, 5], and two, to incorporate improved fabrication and other new techniques in chip fabrication [12, 7, 22], so as to decrease both the Josephson inductance and shunt capacitance in addition to minimizing the material losses. The details of the implementation of this strategy are described as follows.

### 3.3. The most optimal transmon material for coherence and scalability

The architecture presented here is based upon tantalum (Ta) as the material of choice to manufacture films for our transmon qubits, i.e. the underlying Josephson junctions (and, if needed, to fabricate SQUIDs for flux-tunable gates), and their coupled resonators, for frequencies up to 19GHz, beyond which we aim at using niobium. Tantalum has come up as a better and low-loss alternative in terms of coherence times, as compared to the conventional niobium and aluminum oxide, owing to its intrinsic stability and low TLS errors.

As shown in earlier studies, it could be done in various ways, mainly by wet etching, but we have determined that for a viable operation up to the operating frequency of 11.3GHz. in the initial trials, this has to be achieved with the help of a dry etching process, as based upon the recent work by Wang *et al.* [12], who demonstrated that qubits prepared with the help of tantalum films incorporating dry etching are possibly better than other material and processes for achieving complex and scalable quantum platforms. However, there are no studies reported so far that would have highlighted the usage of tantalum in production of high-frequency qubits. In that case, for frequencies beyond 10GHz, the alternative material of choice is niobium (Nb) for our prospective range. Therefore, with the help of niobium films, and adopting approaches such as the design of a tri-layer structure, one could possibly achieve a higher frequency operation and venture up to frequencies of more than 11.3GHz and up, possibly up to 75GHz (or even beyond to the millimeter wavelengths, the so-called TeraHertz regime.)

The successful exploration of stable tantalum transmon units was first demonstrated by a Princeton-based group, reporting a device with a long coherence time of around 0.3ms (300□s) [8], employing a wet etching process. However, exploring the potential of dry etching, which has obviously great benefits over its wet counterpart, Wang *et al.* [12] in 2022, reported longer coherence times of up to 503□s with transmon qubits using Ta films. Recently, a breakthrough has been achieved in this regard, once again by a group at Princeton, Bland and Bahrami *et al.* [7], who reported qubit lifetimes and coherence times of up to 1.68ms (with time-averaged quality factors exceeding $1.5 \times 10^7$) while achieving a gate fidelity of around 99.994% with tantalum transmon qubits fabricated on high-resistivity silicon substrates. In addition, another recent study carried out at Aalto university [22] has demonstrated a similar result reporting their design of a high-coherence qubit with unprecedented energy relaxation, with coherence time surpassing 1.0ms threshold. It appears that tantalum on high-resistivity silicon substrate (Ta-on-hi-res-Si) platforms (involving silicon resistivities on the order of $\geq 20k\Omega$) offer longer qubit lifetimes ($\geq 1.6ms$) and high gate fidelities ($\sim 5 - 7 \times 10^{-5}$) as compared to their tantalum on sapphire or other substrates, in addition to offering easier microfabrication and improved scalability. The reason for this advantage, as observed by Bland and Bahrami *et al.* [7], is mainly lower dielectric losses with the use of high-resistivity Si substrate (reducing both the surface and bulk substrate losses), thus, in turn, achieving higher quality factors. Hence, tantalum on a high-resistivity silicon substrate has clearly great advantages for high-coherence, scalable quantum computing.

There are some other novel materials that have potential for making transmon qubits other than the conventional superconductors, some of these include a class of materials known as the Topological Insulators (TI's). Recently, there have been studies assessing the application of these materials in manufacturing transmons. One of these studies, reported earlier this year, involves making 3D transmon qubits with S-TI-S Josephson junctions based on the three-dimensional TI material $BiSbTeSe_2$ (BSTS) [23]. The particular TI alloy the team employed was $(Bi_{0.06}Sb_{0.94})_2Te_3$, the samples of which were grown with the help of area-selective molecular beam epitaxy, incorporating 100nm wide and 20nm thick TI nanoribbons, and after the fabrication they were complemented by superconducting niobium stencil contacts.

### *3.4. Higher Q value and coherence with optimized capacitor design*
Capacitance holds a key eminence in a successful transmon design, playing a central role in both operating frequency and quality value, thus in turn, determining the qubit's salient characteristics and coherence.

We plan to achieve a significant throughput in our qubits' coherence with the help of trilayer Nb or Al-AlOx-Al high-quality-factor, low-loss and compact ($< 100\mu m$ pads) capacitors for high-freq. ($> 10GHz$) transmon design. These capacitors offer exceptionally low dissipation at microwave frequencies, while offering a small footprint and parallel capacitor geometry, in addition to offering demonstrated high specific capacitance (typically around $12fF/\mu m^2$) and low-loss tangent (on the order of $10^{-6}$) and high reproducibility ($< 2\%/2''$ wafer) [24], the factors which greatly help in facilitating long relaxation times and high coherence in addition to reproducibility. In addition, it could be helpful to employ (in association with planar transmon arrays) various conducive capacitor geometries, such as utilizing a double pad capacitor geometry with shape optimization (e.g. optimized spline-based shapes for the pad and wire, with the former smoothly tapered near the junction wire and balanced properly), an approach that demonstrated a reduction in dielectric losses in the capacitor pads by around 16%, and in turn increased the TLS-limited Q factor and qubit relaxation time by more than 20% [25], or another recent approach, whereby parallel-plate capacitor like geometries involved paddle capacitor arms across the junctions, separated by several tens of micrometers on a silicon or sapphire substrate as demonstrated in recent studies [26].

### *3.5. Transmon fabrication and integration*
The final and most crucial step in the road toward high-coherence, high-frequency quantum computing is the fabrication of actual qubits on their wafers along with their resonators, forming wire contacts, and the packaging of the chip, followed by the placement and operation of the device on the fingers of a dilution refrigerator.

The Josephson junctions are manufactured with the material of choice being tantalum/niobium (Ta/Nb) or aluminum/aluminum oxide (Al/AlO$x$), the former with the help of sputtering while heating at around 500-600°C temperatures, such as described in the recipe by [8], and the latter by means of the two-step process of electron-beam lithography and double-angled aluminum deposition [4]. At first, the suitable high-resistivity silicon wafers with the desired specifications are obtained and prepared, followed by deposition of Ta/Nb films on the substrate with sputtering while heating and proper cleaning of the samples. Various base components of the transmon, mainly the resonators, coplanar waveguides, the ground plane and capacitors are patterned in this way in a step by step manner. Once the base is prepared and properly coated and a resist mask is prepared for placing the junctions (as described in detail in [22], [28], etc.), the step of junction deposition begins involving deposition of the layers and contacts of the Josephson junctions, in a procedure outlined in [22]. In addition, as shown by Anferov *et al.* [4, 5] for their millimeter-wave qubits, and elsewhere, our aim is to achieve a trilayer, double-sandwich Nb/Al/$AlO_x$/Al/Nb Josephson junction structure. Moreover, as shown by Bland and Bahrami *et al.*, by successfully reducing hydrocarbon contamination while depositing and oxidizing the Al/AlOx during the manufacturing process making use of special conditions, the coherence of the devices (relaxation times as well as Hahn echo coherence times) can be improved manifold.

In order to achieve a high-coherence operation as desired, we find it helpful to incorporate some ideas and techniques in fabrication as suggested in detail elsewhere by various investigators, mainly Tuokkola *et al.* [22], Place *et al.* [8], Wang *et al.* [12], and Gingras *et al.* [27] in their reports. All of these reports advocate carefully devised fabrication, chemical cleaning (and in one report [12] also chemical annealing for reducing Material Substrate, MS, and Material-Material, MM, interface losses) procedures to remove oxide and other layers, and proper definition of the capacitor and resonator sections in the device, among other things, for achieving long coherence times. In particular, Gingras *et al.* have pointed out the importance of post-fabrication conditioning in addition to the necessary steps in fabrication. In their recent report [27] they have shown that a surface treatment in the form of a fluorine-based application aided with argon-based ion milling helps remove the germanium and oxide residues, resulting in conspicuously improved relaxation times (with a median of 334 μs and a corresponding Q value of 6.6 x $10^6$). It is also imperative that proper characterization of the tantalum (or Nb/Al/AlOx) films be carried out after fabrication, with the help of appropriate STEM microscopy and spectroscopy, whatever applicable and available, as delineated in [8].

Following the approach by Anferov *et al.* [5], the Josephson junction, its accompanying bias capacitor, ground plane and the readout resonator are all designed and fabricated as integrated on the same plane. The qubits are designed in a grounded configuration but a better approach, though more involved, can be to use them in a differential configuration, achieving an enhanced operation.

The coherence of qubits also depends upon a number of other factors, such as an intrinsic time-dependence, sensitivity to the thermal environment, and quasiparticle generation within the qubit junctions and pair breakage owing to high-energy particle impacts from the environment (such as cosmic ray photons, etc.) [3, 4, 5, 6]. Thus, careful modelling of the qubit temporal and thermal behavior as well as astute electromagnetic and cosmic ray shielding considerations are mandatory.

In terms of choosing the device dimensions, the decisive factors are the frequency in consideration, optimizing the coherence, gate fidelity and coupling, with the main priority on designing the sub-micron Josephson junction and the macroscopic capacitor pads (and the dielectric between them), as they form the core of the device. It is hoped to achieve optimal finger dimensions within the range 0.6-0.8 μm [12]. However, we choose the 0.65 μm sweet spot for achieving the qubit center frequency of interest in this case. Each arm of the transmon is planned to be around 300-350 μm long and is to be optimized to form an ideal capacitance of around 47fF.

### ***3.6. The transmon coupler geometry and a multiple transmon-resonator qubit architecture***

A useful and scalable quantum computing architecture involves many qubits which are obtained by useful geometries of multiple transmons coupled with their associated couplers and resonators connected to the respective data and control buses, so that control and readout protocols could be implemented.

Architectures involving two and three transmons connected to a single resonator (via respective couplers) have been reported in literature and have been found to be quite effective in increasing system scalability. In particular, there are various studies that involve two transmons associated with a single resonator, i.e. the Double-Transmon-Coupler (DTC) scheme [16, 17], and a three-transmon architecture [18], where these ideas were originally developed. There are also some proposals for four-body and higher architectures reported in the literature [29].

The qubits in the current design are designed in terms of a planar geometry in sets of four transmon qubits with corresponding X-shaped coupling capacitors to a single quarter-wave resonator, coupled capacitively to the former, and the latter is, in turn, connected via inductive coupling to a transmission line, a part of a feedline backbone for measurement and readout. Thus, a four-body, Quad-Transmon-Coupler (QTC) architecture is obtained. The two quad-transmon coupler units achieved in this way are then connected to a common Quantum bus. There are separate lines in the design for flux biasing of the transmons, if a flux-bias operation is entailed.

In the design conceived here, a set of four capacitively coupled one-quarter (λ/4), or alternatively one-half wavelength (λ/2), readout coplanar waveguide resonators, having frequency of approximately 18GHz, are incorporated for each of the qubits, which have a central resonance frequency ($\omega_q$) of 11.3GHz (or as desired). Based upon our own design priorities as well as following the guidelines by Kurilovich *et al.* [32], we keep the readout resonator frequency much higher than the qubit frequency in order to increase coherence time as well as suppress spurious resonances. Purcell filters are introduced at the beginning of each feedline owing to our working in strong transmon-coupler coupling regime, limiting one filter to at most four (ideally three) resonator-qubit loads, as illustrated in the Figure 4, an overview of an 8-qubit high-frequency system design proposed in this work. Thus, in this way, the mix of a centralized as well as a distributed architecture design is obtained. Figure 3 illustrates an overview of a proposal of the simple Quad-Transmon-Coupler (QTC) architecture, presented here. The transmons are depicted by a "Q", such that four transmons are connected via their respective couplers (depicted by a symbol of capacitor) to one resonator (depicted by an orange square in the center.) There can be many variations of this simple topology where the number of transmons, as well as their connectivity, could be altered.

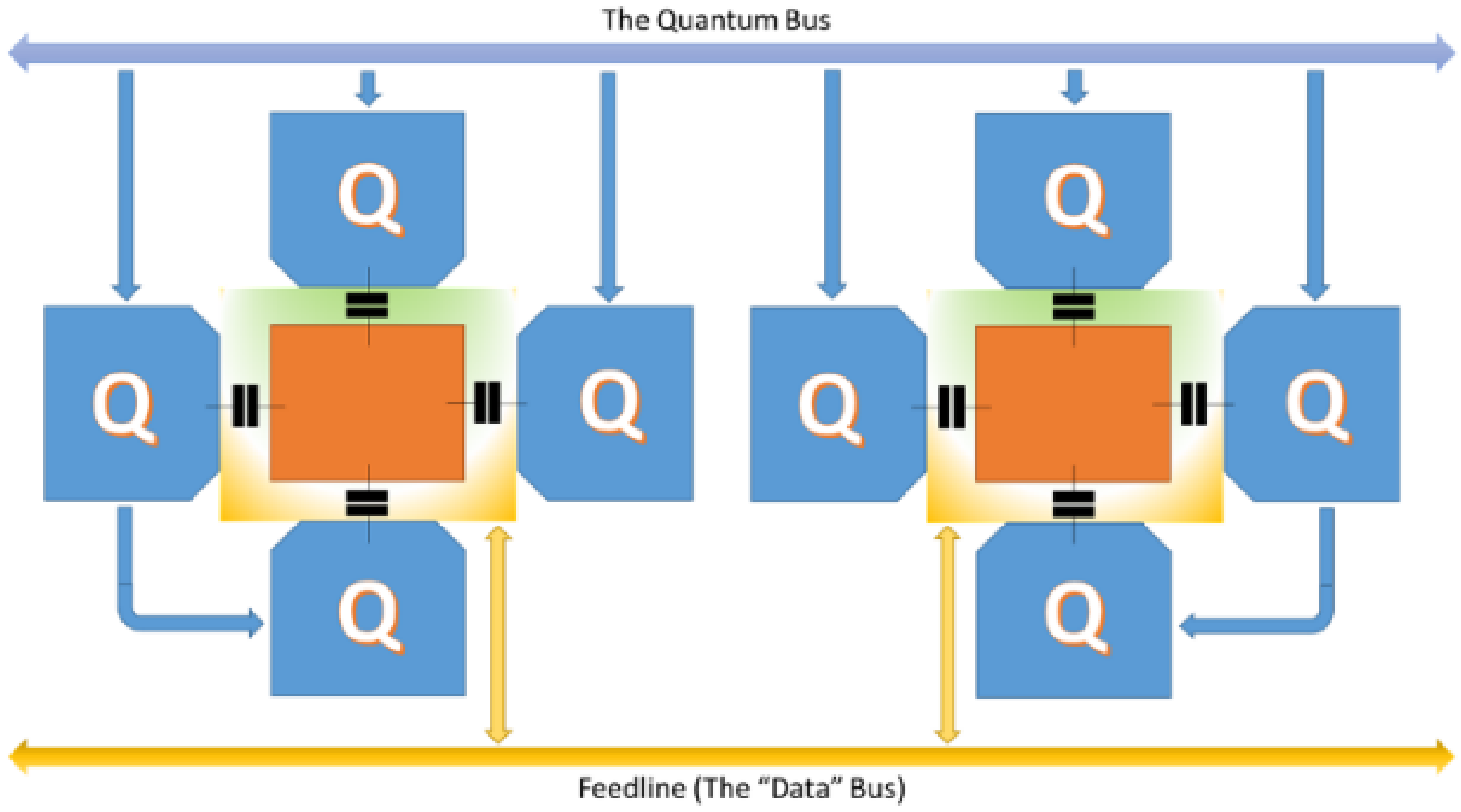

FIGURE 3. A Schematic of the simple Quad-Transmon-Coupler (QTC) topology qubit architecture proposed here. Four qubits (depicted by blocks in blue color) are connected to one resonator (the block in orange color) via couplers.

Figure 4 illustrates an overview of our proposed 8-transmon-coupler-resonator architecture on a chip. The geometry and overall design of the transmons and their associated couplers are also illustrated in the Figure 4. The figure also illustrates an overview and dimensions of a sample transmon in this case (the design and dimensions can vary owing to various factors.) The Figure 9 (Figure SM1 in supplementary material) presents an overview of an earlier low-frequency (4-6GHz) and flux-tunable version of the architecture conceived by the author. However, the version reported here is the high-frequency and finalized version of that, using Niobium. The design parameters for the qubit and resonator were obtained by relevant calculations as well as with the help of simulations carried out using the ANSYS package.

The important parameters of the proposed architecture and the particular design targets conceived here (based upon our calculations and simulations) are as enumerated in Table 1.

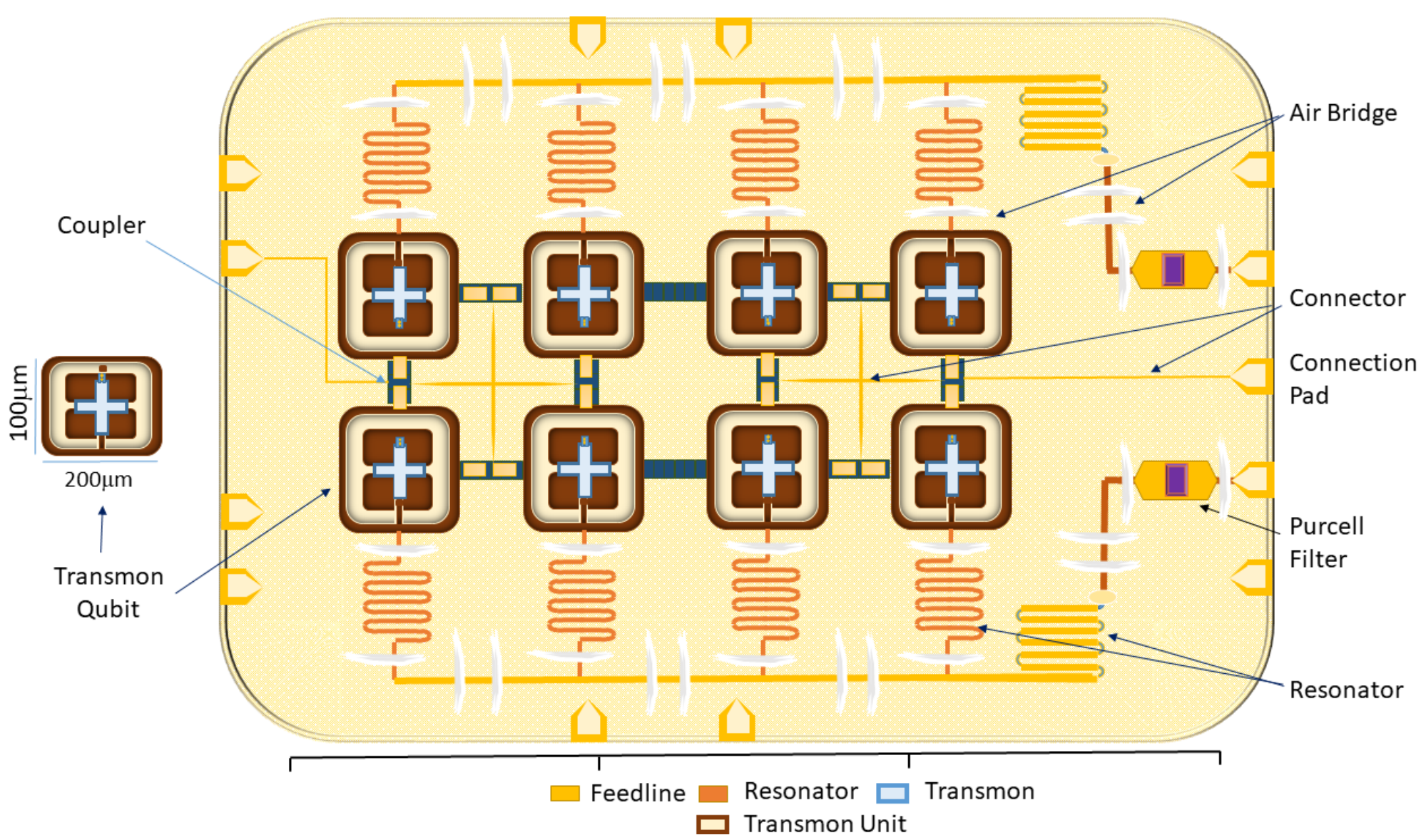

Figure 4. The Eight-Transmon Qubit Architecture Design with Couplers, Resonators and Feedlines—An Overview Diagram of the High-Frequency 8-bit Quantum System Architecture Presented Here. A Single Transmon Unit with Junctions and Bias Capacitor Pads is Shown on the Left.

## 4. Readout and Control

The three integral components of a quantum computing experiment or system are the qubit, the readout and control system (including the digital computing section, which is achieved with the help of a conventional digital computing device) and the cooling and enclosure assembly system (including shielding). Once a transmon-based qubit device and resonator assembly is designed and developed, the second section is designed and put in place, followed by the design and development of the cooling and enclosure system and putting in place a proper shielding.

### *4.1. Precision, low-level quantum signal radiometry*

The readout of a transmon qubit is in its nature a quantum weak signal measurement and readout, and is not different from conventional radio frequency radiometry, since the first proper radiometry device was innovated by the genius of Robert Dicke during the 60s [31, 32], however, it is quite involved, since the amplitude of the signal from a transmon qubit is very weak and involves many challenges, as documented in detail in our earlier report on the subject [35]. Modern weak photon radiometers, especially the ones operating at high frequencies, involve a chain of sensitive low-noise amplification devices operating at various operating temperatures.

The three important sub-sections of this section are the cryogenic readout and control (mainly isolating, filtering and amplification), the room-temperature electronics and the interface with a digital computational and control system.

The readout and control system, as designed for our proposed architecture, are illustrated in the Figures 5 and 6.

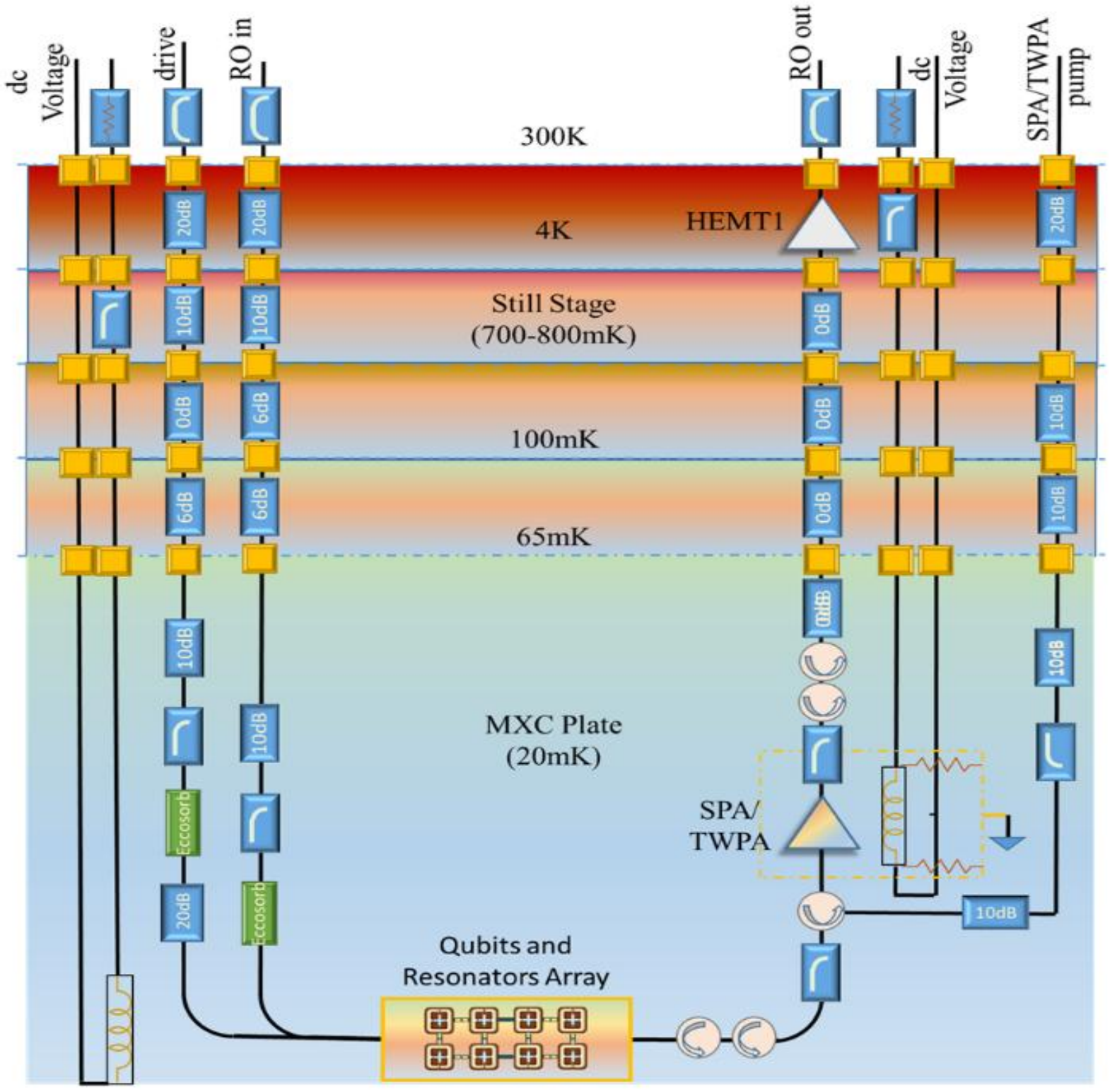

Figure 5. The Experimental Setup Overview: A Schematic of readout and control electronics, including the cryogenic and room-temperature devices and connections. The first stage is a SNAIL Parametric Amplifier/Travelling Wave Parametric Amplifier (SPA/TWPA) with its associated pump and readout circuitry. The sensing and readout chain signals from the transmons are first filtered by HERD IR blockage filters, followed by circulators and low-pass filters and attenuators in order to achieve the required envelope. The drive chain has a similar HERD filter at the MXC plate followed by a series of filters and attenuators. All signal lines have band-pass filters at the respective ports.

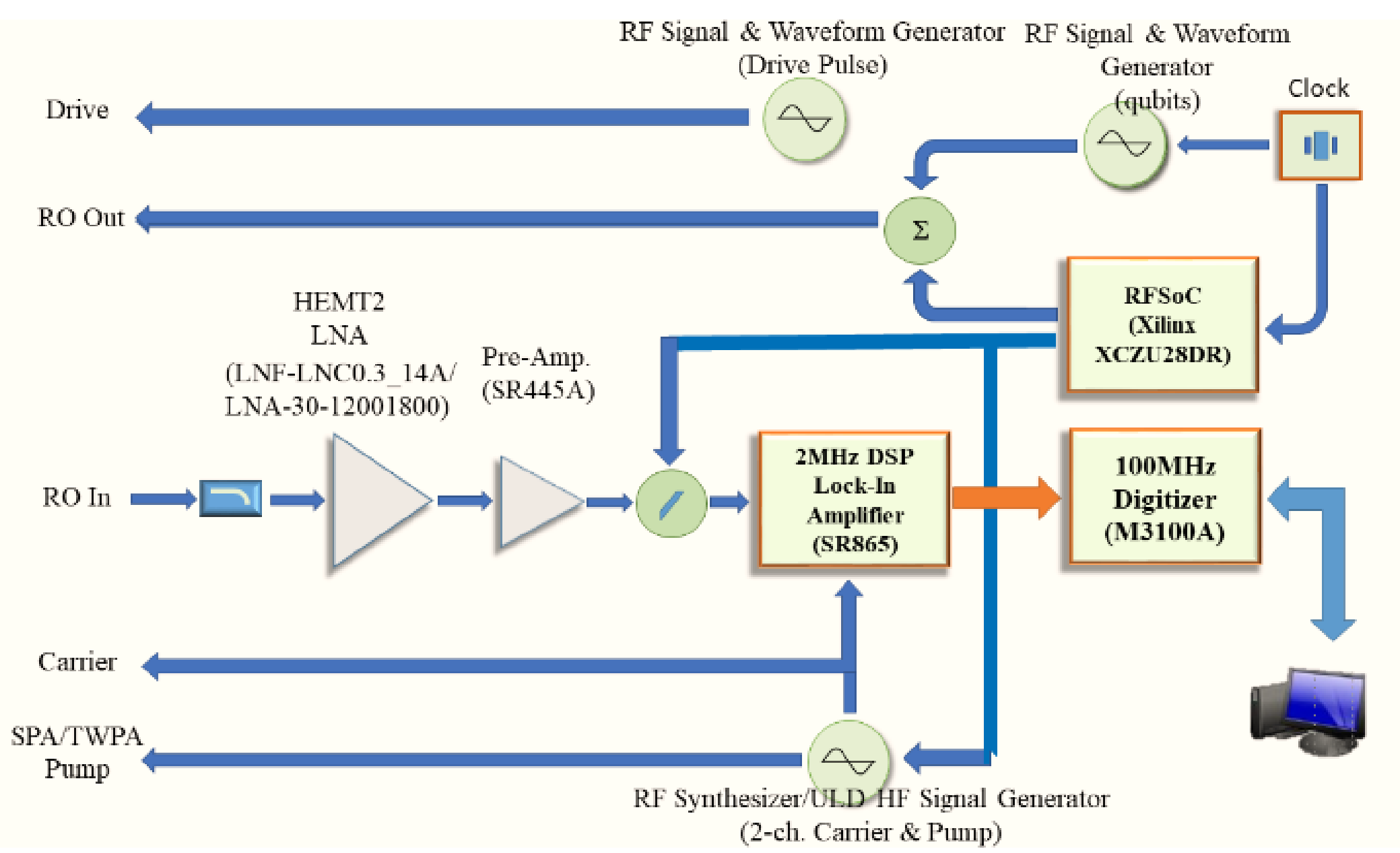

Figure 6. A Schematic of the Readout and Control Electronics, the Central Part of the Scheme is a Xilinx Corp. RF Systems on a Chip (RFSoC), which in our implementation is an XCZU28DR Evaluation board, based upon the Xilinx Field Programmable Gate Array (FPGA) chip, the ZCU111.

### *4.2. TWPA/SNAIL-based amplification*

The main amplification device in the first (and crucial-most) signal readout stage in the current design, following an array of high-frequency (12GHz) circulators/isolators and filter(s) acquiring the signal from the qubits and resonators array chip, is a Travelling Wave Parametric Amplifier (TWPA) [10, 31], a superconducting parametric amplifier made with an array of SQUIDs (Superconducting Quantum Interference Devices) or JPAs. In this context, we plan to implement a recent innovation in the device, a Superconducting Nonlinear Asymmetric Inductive eLements (SNAIL) array-based parametric amplifier (SPA) [53, 33], which is a quantum-limited Josephson-junction-based 3-wave-mixing parametric amplifier device, and in the essence, a long chain of superconducting non-linear amplification elements [32, 34] — a new and recent feature improved from our earlier designs involving a simple Josephson Amplifier (JPA), which was suggested in our earlier studies, especially in our reports [9, 35]. The device needs to be run in a phase-sensitive mode so as a measurement of up to at least 85-90% could be achieved [32].

The TWPA/SNAIL device is pumped by a TWPA pump line obtained from a signal generated by a commercial RF synthesizer (Maury Microwave-Holzworth HS9100 four-channel phase-coherent RF synthesizer). In view of earlier experience, we plan to pump the amplifier with an on-chip pump current of around 90nA to achieve the maximum possible gain (of around -12dB to -15dB) and pump power (of approximately -90dBm to -93dBm) at the pump frequency. The output of the TWPA/SNAIL stage is further amplified by a long amplification chain, involving at first a cryogenic High-Electron Mobility Transistor (HEMT) LNA device (a Low-Noise Factory/Narda-Miteq LNA, LNF-LNC0.3_14A, operating at 4.0K, which was found to be the lowest noise factor and highest gain device in our survey, more details on it in the reference/footnote [36]), followed by another LNA, the HEMT2 (a NARDA-MITEQ LNA part number 30-12001800, a device offering a high-gain of around 30-33dB and a low-noise figure of around 1.3dB, and a gain flatness of $\pm$1.5dB, operating at room-temperature). This dual combo stage increases the strength of the weak signal detected from the qubit system to a sufficient stage to be readout by instruments. For optimal signal measurement, there are low-pass, high-pass and band-pass filters installed in the amplification chain, in their respective thermal environments. The readout and drive lines are attenuated where necessary by suitable attenuators as illustrated in the figure 5, whereas appropriate signal filtering is achieved by special "Eccosorb" (Laird Technologies), polyutherane foam-based EMI filters, and where necessary by other commercial filters. We aim to install new, specialized non-magnetic, in-line High-Energy Radiation Drain (HERD) carbon low-pass filters in both the drive and readout chains at the MXC stage, in view of the IR radiation flux that could be detrimental to coherence, in particular the HERD-2 IR blockage devices (Quantum Microwave, inc.) with excellent low-loss and in addition high stop-band attenuation up to 13.5GHz, ideal for our application.

An overview of the whole cryogenic readout and control system (excluding the external instrumentation) is illustrated in the Figure 5, with fictitious color coding depicting varying degrees of operating temperature in various stages.

The next stage, i.e. the readout (and control) instrumentation, could be any of a homodyne, heterodyne/superheteodyne or spectral analysis mode signal detection, one could choose any one of those methods of signal readout, depending upon various factors such as system complexity and cost. When the cost is not a concern, an off-the-shelf commercial turn-key solution, such as a Zurich Instruments Super High Frequency Qubit Controller (SHFQC) instrument, could be utilized that performs all the entailed readout and control operations of the qubits (not tested or verified by us.) Alternatively, a modern RF Systems on a Chip (RFSoC), with its respective FPGA digital circuitry as well as ADCs (Analog to Digital Converters) and DACs (Digital to Analog Converters) and a rack of precision instruments in a laboratory can do the tasks.

***4.3. Homodyne/heterodyne/super-heterodyne signal readout***

In the current implementation of our design, we choose the mode devised in our earlier studies as reported in the literature [9] and the references of our earlier works therein. We utilize a commercial low-noise, high-sensitivity pre-amplifier (a Stanford Research SR445A 4-channel, dc to 350MHz low-noise (6.4nV/√Hz) amplifier) and a lock-in amplifier (Stanford Research SR865 2MHz LIA). The amplified high-frequency GHz signal is mixed with a carrier signal from a local oscillator (LO), in our case fed by the RFSoC, and amplified by the Lock-in Amplifier (a Stanford Research SR865 2MHz DSP LIA), the output of which is digitized by a commercial analog/digital converter (a Keysight M3100A 14-bit, 100MSa/s digitizer) and readout by a computer system. For simplicity, the LIA stage could be removed and the signal can be mixed with the carrier signal and fed directly to the digitizer, however, it would reduce the signal to noise ratio (SNR).

**Table I. Salient Design Parameters.**

| | |
|---|---|
| Number of Qubits on-Chip as of now | 8 |
| Configuration Technology | Transmon |
| Geometry | Coplanar |
| Superconducting Junction Material | Tantalum (up to 11.399 GHz as of now) Niobium/Aluminum/TiN (up to 20-75GHz and possibly beyond) |
| Substrate material | High-resistivity Silicon / Sapphire |
| Coupler Geometry | Quad Transmon |
| Pitch | 0.5-0.75 μm |
| Chip Dimensions | 5mm x 5mm |
| Etching Process | Dry Etching |
| Optimal Thermal Stability | 50-75mK. 65mK (mean); 100 mK (max.) |
| $\omega_1/2\pi$, $\omega_r/2\,\pi$ | 11.3 GHz, 11.99 GHz., 17.99 – 19.27GHz. |
| $g/2\pi$ | ~ 109.2 MHz. |
| $\alpha/2\pi$ | 380.7 MHz. |
| Avg. Exp. Q value $\langle Q \rangle$ | $1.75 - 2.75\ \times 10^7$ |
| $\langle T_1 \rangle$ $\langle T_2 \rangle$ | 0.2 – 0.3 ms 0.4 – 0.6 ms |
| $E_J/2\pi$ | 41.37 GHz |
| $E_C/2\pi$ | 299 MHz |
| $E_J/E_C$ | 139 |
| $L_J$ | 2.75nH |
| $C_\Sigma$ | ~ 37.0 – 58.3 fF |
| $I_C$ | 122.75 - 139.0 nA |
| $J_C$ | 0.21-0.39 $kA/cm^2$ |
| $R$ | $\sim 2K\Omega$ |
| T | 0.02 – 0.065 K |
| Mean Quant. Meas. Eff. <η> | ~ 0.0002799 @ $T_{sys}$ = 0.020K. |

The mastermind for the whole experiment/platform in our particular design is an RF Systems on a Chip (RFSoC) from Xilinx Corp., an XCZU28DR Evaluation board, based upon the Xilinx Field Programmable Gate Array (FPGA) chip, the ZCU111 [37]. There are some recent versions of the controller available in the market now, such as the AMD-Xilinx "UltraScale+ RFSoC Gen. 3" series, such as the ZU49DR chip, available also in the evaluation board form (the part number ZCU216).

In addition to the instrumentation shown in the diagram, it is essential to carry out spectral analysis of the obtained high-frequency signals from the TWPA/SPA, the cryogenic HEMT amplifier and also from the room-temperature amplifier, in connection with testing and calibration of the involved signals and methods. In the absence of an expensive high-frequency Vector Analyzer, this feat could be achieved with the help of the HS9100 source and a suitable spectrum/RF signal analyzer, such as Rhode & Schwarz FPS 13.6GHz Signal & Spectrum Analyzer (for high-frequency) or an Aaronia GmBH Spectran V8 Plus 8 GHz Spectrum Analyzer (for low-frequency).

Similar to the study carried out by Ding *et al.* [38], all the *ac* signals required for qubit driving, control and readout are to be provided by commercial RF generators (1GSa/s arbitrary waveform generators, such as the Keysight M3202A, in combination with Rhode and Schwarz SGS100A SGMA or HS9100 RF sources or whatever available). The LIA lock-in amplification is done in association with a carrier signal generated by an RF synthesizer/RF source or an Ultra-Low Distortion High-Frequency Signal generator with low-noise ~1MHz operation. We suggest using one of the channels from the HS9100 RF synthesizer or a separate low-cost ultra-low-distortion RF source (Stanford Research ULD signal generator [35]) to generate the carrier signal for LIA. A 10MHz rubidium crystal time base is used to provide the master timing clock for all the required synthesis, digitization and control synchronization.

More details on the RF readout, RFSoC and associated technology can be found in the literature, such as in the references [39-43].

### *4.4. The cryogenics and shielding*

The central non-intelligent part of the whole experiment/system is a Bluefors XLD600 (operating at 20mK) dilution refrigerator (unless $^4$He adsorption-based cooling is used), however, any modern dilution refrigerator with capabilities to reach down to the temperatures of 20mK may be used. Owing to the influx of thermal photons and consequently going out of thermal equilibrium of the qubits at low temperatures due to coupling to the readout chain and the environment, it is expected that the current design could operate at ambient temperatures of around 30mK, at minimum, to 50mK, at maximum, depending upon qubit fabrication and architecture implementation. Figure 7 illustrates an overview of the experiment mounted on the cold plates of a dilution refrigerator, where the qubit arrays, encased in their assembly can, are mounted on the Mixing Plate/Chamber (shown as MXC in the figure). The cold plates mainly used, other than the MXC/MC are the "Still" (at 100mK to 400mK temperatures) and the 4K stages.

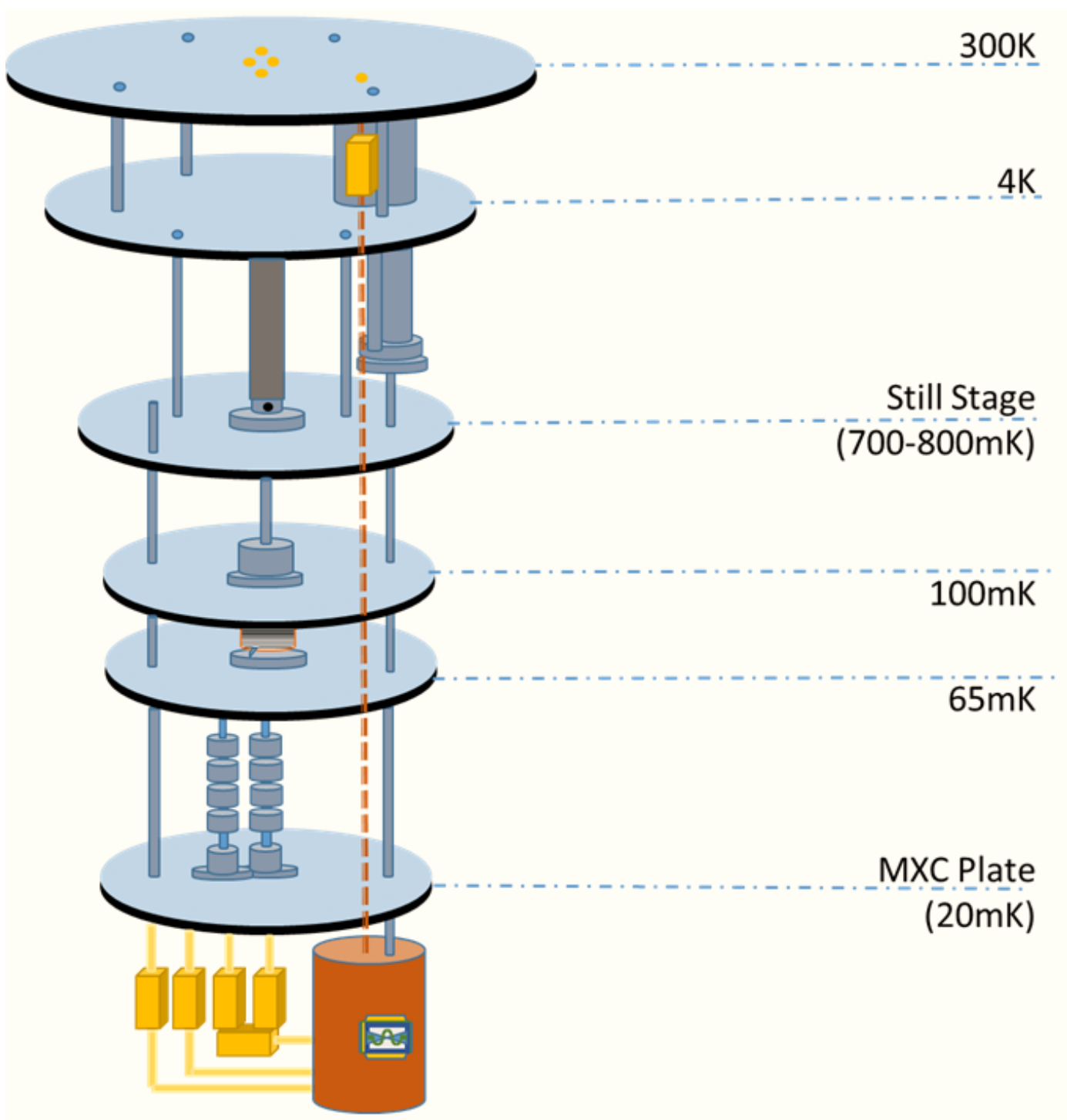

Figure 7. The Complete Measurement Setup on Mounted on a Dilution Refrigerator (the drawing of the dilution refrigerator is inspired by the work by Cai *et al.* [46]).

The alternative choice other than employing an extremely expensive dilution refrigerator, as usual, is using the new method of liquid Helium-4 ($^4$He) adsorption-based refrigeration, as demonstrated recently elsewhere [5]. This technique has immense potential for achieving high-frequency quantum computing operation in small laboratory settings, if correctly implemented.

In terms of fabrication, packaging and three-dimensional integration of this or any high-frequency high-coherence qubit design the main concerns should be reducing the parasitic two-level system losses among other things. Figure 8 illustrates an overview of the prospective chip connections and chip packaging as well as the shielding "can" assembly in which the former, i.e. the chip, is mounted laterally on the mixing plate. As shown in the figure there would be multiple layers of appropriate shields in place, to reduce the experiment's coupling to nearly all kinds of external interferences, mainly thermal, magnetic, electromagnetic and cosmic and other possible radiation.

More details on appropriate microwave chip package design could be found elsewhere, such as in the reference [44]. There would be concerns about wiring and interconnects, which would entail impedance-matched high-frequency (10-20GHz) coaxial wiring as well as connectors for all signal transmission, readout and control. In addition, in order to protect our flux-tuned qubits from external electromagnetic interference from the environment as well as from ambient magnetic fields, a tight, multi-layer and suitable shielding would be extremely paramount, including commercial mu-metal magnetic shield, such as "Amumetal" or "Cryoperm" [45] shields, and "Berkeley Black" or "Vantablack" commercial IR shields.

As is the norm, the whole experiment has to be installed in a special superconducting can enclosure assembly with necessary ambient magnetic field, flux noise and cosmic radiation shields [22] (at least the three shields, viz. the inner IR and superconducting shields combo, the mu-metal shield, and the outermost lead shield) before mounting on the mixing chamber of the dilution refrigerator.

The chip has to be fabricated on an especially-devised low-loss K-band circuit board. Finally, the chip would be wire-bonded in a typical commercial chip carrier (such as a QCage.24 carrier) and connections would be taken out. The chip package is mounted vertically in the assembly can before installing it in the dilution refrigerator/cryogenic stage.

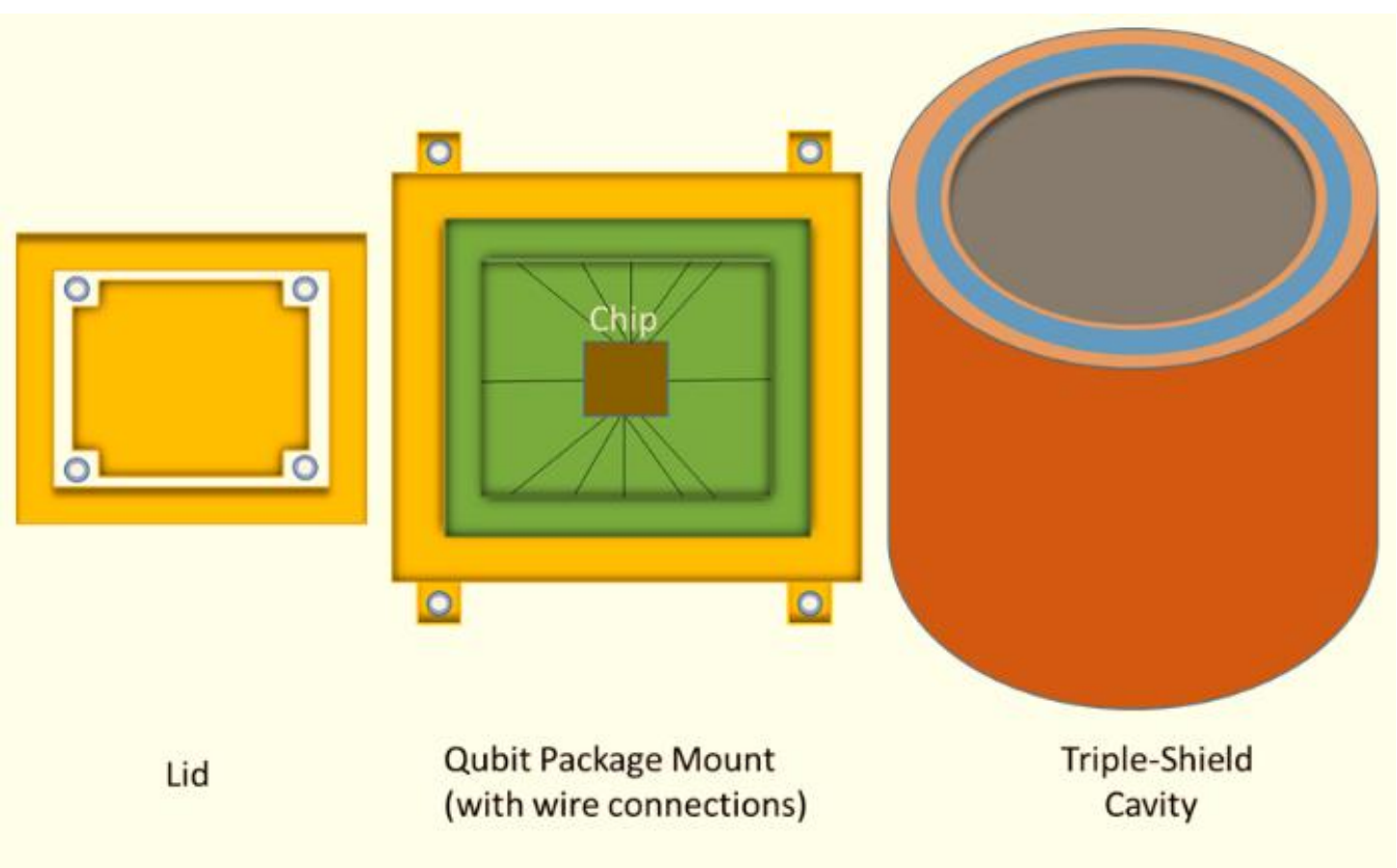

Figure 8. An Overview of Chip Connections and Packaging as well as the Triple-Shield Cavity Shielding Assembly (the "Experiment Can"). The Qubit array is shown to be packaged in its package in the center with its Lid removed (shown on the left), whereas the three-shield can for the package is shown on the right.

Thus, in this manner a Quantum Processing Unit (QPU) is achieved that is installed on a dilution refrigerator or a suitable alternative low-temperature mount and is connected to the control and readout circuitry. Necessary details on the operation and usage of a QPU in association with conventional computing, by means of carefully devised algorithms (including shaping envelopes etc.), and fed into the device with the help of microwave pulses may be found in the relevant literature, such as in [32].

## 5. Conclusion

We present in this report the design of a high-frequency transmon-based qubit system geared towards high-coherence quantum computing, along with preliminary designs of the involved key components. Besides, some guidelines are also presented for the optimal working of the designs suggested here. The most crucial elements of a viable high-frequency quantum computing architecture are a robust transmon design with a sustained operation, long coherence times and low gate errors, coupled with scalability. In order to implement such an architecture design, the most important factors are the choice of superconducting junction material, topology, coupler and resonator interface and successful low-noise readout and external control (with a least coupling to the environment.) Recently, there have been some encouraging numerical studies reported in the literature whereby high-frequency operation of multiple qubits has been computationally achieved at the frequencies of interest (up to 13GHz) [47].

The qubit coherence losses are the most important sources of decoherence and a great challenge, and in this context, the main steps taken to minimize these are reducing material defects, the TLS defects in the material interface, processing artifacts and unwanted depositions, suppressing charge and flux noise, and incorporating transmon and coupler elements and topologies that could be conducive to increased coherence. The other factors that are important in the successful implementation of a quantum computing architecture are a bit higher than the conventional high-temperature operation (beyond the current 65mK limit) and a small footprint. Rest virtually everything lies in the domain of algorithms, i.e., the software.

Moreover, with the help of carefully devised algorithms and FPGA decoder integrated into the qubit control system, as demonstrated elsewhere in several studies, such as by Huber et al. [52], it is hoped to achieve low-latency and at the same time a real-time error correction.

It is planned to incorporate these ideas and the preliminary design (that was an independent effort) into a well-coordinated and concerted, collaborative effort with colleagues and institutions with relevant fabrication facilities and expertise available, and is hoped that in due course (scalable and stable) an optimal, high-coherence and high-frequency quantum computing platform could be achieved. The aim is to minimize in the fabrication and development process all the three crucial factors of the qubit-environment coupling, the control pulse distortion and the magnitude of disruptive heat from control pulses. This is extremely important.

However, there is still space for improvements in the current design that could be taken care of in future designs, especially for a qubit number of more than 100 units and endeavoring to operate the device in the temperatures above the current design's upper limit of around 100mK (the qubits' actual current operating range at the moment is expected to be limited to around 50-65mK).

It is hoped that the ideas and design considerations presented here could be conducive to achieving viable high-frequency, high-coherence transmon-based quantum architectures, and with necessary improvements could become the base of a working model. A successful model should be able to fulfill minimum requirements entailed for a working quantum computer, such as those suggested by David DiVincenzo (the so-called "DiVincenzo Criteria").

The first aim of the design presented here (and of any quantum computing architecture design, in general) should be at first a successful implementation of a suitable quantum gate, such as a Toffoli gate. Once a Toffoli gate is implemented and a sufficient coherence and computing fidelity are achieved, other computational algorithms could be implemented, harnessing the architecture's full capabilities.

There are countless viable applications of the architecture presented herein and a couple of plausible approaches to optimize the working of design proposed here, one of those is using a Reinforcement learning-based control of the operation of a quantum computing device, as recently proposed by a team at Google Quantum AI [48]. It has been demonstrated experimentally that RL-based methods help achieve much higher performance and error correction throughput as compared to conventional error correction methods. A great advantage of this approach is in scalability- the research has shown that an RL agent-environment system can be scaled to much larger and complex systems, which are endowed with optimization speeds that are especially independent of the system scale. With the help of this approach, an agent is trained with the help of Reinforcement Learning (RL) [49] to make suitable countermeasures to preserve the state of a qubit (and hence the information contained within it), based upon a feedback from measurements of a current state of a qubit, in effect minimizing the errors and the RL agent working as an intelligent Quantum Error Correction (QEC) system. The team employed single-qubit gates in the form of shaped microwave pulses (with well-defined in-phase and quadrature components) using a "Derivative Removal by Adiabatic Gate (DRAG)" correction approach [50] to minimize the qubit leakage to higher-energy states. The new RL agent, named Agent-R1 framework [51], is based upon extended Markov Decision Process (MDP), especially aimed at training Large Language Models (LLMs) for complex talks. A successful implementation of this architecture could be utilized in evaluating such LLM-based models.

Therefore, in the end, it is pertinent to assert that in charting the most viable roadmap to efficient and productive quantum computing there are a number of ways that could possibly be reached with continuous testing, improvement and with productive collaboration between research teams.

**Acknowledgments**

This research is an independent effort and did not receive any support or funding, however, the author acknowledges the faculty support and research funding support for his laboratory by the Department of Physics, Jazan University, and from a grant under the Deanship of Scientific Research, Jazan University, and a competitive research grant from the KACST, Saudi Arabia, in the past. Author acknowledges the valuable local support by Physics Division, U. S. Jefferson National Laboratory, Newport News, VA, during visits and stays there during the last eleven years, and thanks their support.

**References:**

[1] A. A. Houck *et al.*, Controlling the Spontaneous Emission of a Superconducting Transmon Qubit, *Phys. Rev. Lett.* **101**, 080502 (2008).
[2] J. Koch, M. Y. Terri, J. Gambetta, A. A. Houck, D. I. Schuster, J. Majer, A. Blais, M. H. Devoret, S. M. Girvin, and R. J. Schoelkopf, Charge-insensitive qubit design derived from the Cooper pair box, *Phys. Rev. A* **76**, 042319 (2007).
[3] R. Barends, J. Kelly, A. Megrant, D. Sank, E. Jeffrey, Y. Chen, Y. Yin, B. Chiaro, J. Mutus, C. Neill, *et al.*, Coherent Josephson qubit suitable for scalable quantum integrated circuits, *Phys. Rev. Lett.* **111**, 080502 (2013).
[4] A. Anferov *et al.*, Superconducting Qubits above 20 GHz Operating over 200 mK, *PRX Quantum* **5**, 030347 (2024).
[5] A. Anferov, F. Wan, S. P. Harvey, J. Simon, and D. I. Schuster, Millimeter-Wave Superconducting Qubit, *PRX Quantum* **6**, 020336 (2025).
[6] P. Magnard, P. Kurpiers, B. Royer, T. Walter, J.-C. Besse, S. Gasparinetti, M. Pechal, J. Heinsoo, S. Storz, A. Blais, and A. Wallraff, Fast and unconditional all-microwave reset of a superconducting qubit, *Phys. Rev. Lett.* **121**, 060502 (2018).
[7] M. P. Bland and F. Bahrami *et al.*, 2D transmons with lifetimes and coherence times exceeding 1 millisecond, arXiv: quant-ph 2503.14798v1 19th March 2025
[8] A. P. M. Place *et al.*, New material platform for superconducting transmon qubits with coherence times exceeding 0.3 milliseconds, *nature comm.* **12**, 1779 (2021).
[9] M. H. S. Bukhari, The search for the cosmological cold dark matter axion – A new refined narrow mass window and detection scheme, *Open Physics* **23**, 20250193 (2025).
[10] V. Gaydamancheko, C. Kissling and Grünhaupt, RF-SQUID-based traveling-wave parametric amplifier with input saturation power of −84 dBm across more than one octave in bandwidth, *Phys. Rev. Applied* 23, 064053 (2025).
[11] D. Face, D. Prober, W. McGrath, and P. Richards, High quality tantalum superconducting tunnel junctions for microwave mixing in the quantum limit, *Appl. Phys. Lett.* **48**, 1098–1100 (1986).
[12] C. Wang *et al.*, Towards practical quantum computers: transmon qubit with a lifetime approaching 0.5 milliseconds, *npj Quantum Inf.* **8**, 3 (2022).

[13] I. Solovykh *et al.*, Few-photon microwave fields for superconducting transmon-based qudit control, *Beilstein J. Nanotech.* 16, 1580 (2025).
[14] V. Tripathi et al., Modelling Low- and High- Frequency Noise in Transmon Qubits with Resource-Efficient Measurement, *PRX Quantum* **5**, 010320 (2024).
[15] C. Han, L. Qi-Chun, Z. Chang-Hao, Z. Ying-Shan, L. Jian-She, and C. Wei, Construction of two-qubit logic gates by transmon qubits in a three-dimensional cavity, *Chinese Physics* **B** 27, 084207 (2018)
[16] H. Goto, Double-Transmon Coupler: Fast Two-Qubit Gate with No Residual Coupling for Highly Detuned Superconducting Qubits, *Phys. Rev. App.* **18**, 034038 (2022).
[17] K. Zhao *et al.*, Microwave-activated high-fidelity three-qubit gate scheme for fixed-frequency superconducting qubits, *Phys. Rev. App.* **24**, 034064 (2025).
[18] S. Ganjam *et al.*, Surpassing millisecond coherence times in on-chip superconducting quantum memories by optimizing materials, processes, and circuit design, arXiv:2308.15539v2 [quant-ph] 14 Sep 2023.
[19] L. I. Glazman and G. Catelani, Bogoliubov quasiparticles in superconducting qubits, *SciPost Phys. Lect. Notes* **31** (2021).
[20] T. E. Roth, R. Ma and W. C. Chew, An Introduction to the Transmon Qubit for Electromagnetic Engineers, arXiv: 2106.11352 [quant-ph] 21 June 2021.
[21] P. Krantz, M. Kjaergaard, F. Yan, T. P. Orlando, S. Gustavsson and W. D. Oliver, A quantum engineer's guide to superconducting qubits, *Appl. Phys. Rev.* **6**, 021318 (2019).
[22] M. Tuokkola *et al.*, Methods to achieve near-millisecond energy relaxation and dephasing times for a superconducting transmon qubit, *nature comm.* **16**, 5421 (2025).
[23] S-W. Huang *et al.*, Towards a hybrid 3D transmon qubit with topological insulator-based Josephson junctions, arXiv:2506.18232v1 [cond-mat.mes-hall] 23 Jun 2025.
[24] K. Silwa *et al.*, Characterization of Transmon Qubits with Low-Loss Parallel Plate Capacitors, APS March Meeting 2023, id. S73.014 (2023).
[25] S. Eun, S. H. Park, K. Seo, K. Choi and S. Hahn, Shape Optimization of Superconducting Transmon Qubit for Low Surface Dielectric Loss, *J. Phys. D: Appl. Phys.* **56**, 505306 (2023) [arXiv:2211.14159 quant-ph]
[26] A. D'Elia *et al.*, Characterization of a Trasmon Qubit in a 3D Cavity for Quantum Machine Learning and Photon Counting, *App. Sci.* **14**, 1478 (2024).
[27] M. A. Gingras *et al.*, Improving Transmon Qubit Performance with Fluorine-based Surface Treatments, ArXiv:2507.08089v1 [quant-ph] 10 July 2025.
[28] S. Kono *et al.*, Mechanically induced correlated errors on superconducting qubits with relaxation times exceeding 0.4 ms, *Nat. Commun.* **15**, 3950 (2024).
[29] R. Miyazaki and T. Yamamoto, Four-body coupler for superconducting qubits based on Josephson parametric oscillators, *Phys. Rev.* **A111**, 062612 (2025).
[30] M. Carroll, S. Rosenblatt, P. Jurcevic, I. Lauer, and A. Kan dala, Dynamics of superconducting qubit relaxation times, *npj quantum inf.* **8**, 1 (2022).
[31] Z. Wang, M. Xu, X. Han, W. Fu, S. Puri, S. M. Girvin, H. X. Tang, S. Shankar, and M. H. Devoret, Quantum microwave radiometry with a superconducting qubit, *Phys. Rev. Lett.* **126**, 180501 (2021)
[32] P. D. Kurilovich *et al.*, High-frequency readout free from transmon multi-excitation resonances, arXiv:2501.09161v1 [quant-ph] 15 Jan 2025.
[33] A. C. C. de Albornoz *et al.*, Oscillatory dissipative tunneling in an asymmetric double-well potential, arXiv:2409.13113v1 [quant-ph] 2024.
[34] N. E. Frattini, V. V. Sivak, A. Lingenfelter, S. Shankar, and M. H. Devoret, Optimizing the Nonlinearity and Dissipation of a SNAIL Parametric Amplifier for Dynamic Range, *Phys. Rev. Applied* **10**, 054020 (2018).
[35] M. H. S. Bukhari, A Table-Top Pilot Experiment for Narrow Mass Light Cold Dark Matter Particle Searches, universe **6**, 28 (2020).

[36] The main LNA device (HEMT1) at the heart of the amplification chain following the TWPA is the LNF-LNC0.3_14A (Low-Noise Factory), the lowest noise possible within this frequency range in the author's survey, offering a 3-14GHz low-temperature operation, with a very low 0.06dB Noise Figure, translating to a roughly 4.2K noise temp., and a 41dB Gain. The LNA chosen (HEMT2) for the next stage, i.e. for further amplification at the room-temperature is an LNA-30-12001800-13-10P (Narda-Miteq) that offers a 12-18GHz operation with a gain of around 30-34 decibels at a noise figure of 1.1-1.3 decibels, however, the same LNA as the cryogenic HEMT, the LNF-LNC0.3_14A, can also be employed for this stage, if price is not a concern.
[37] ZCU111 Evaluation Board User Guide (UG1271), Xilinx.com (AMD-Xilinx, Oct. 2nd, 2018).
[38] L. Ding *et al.*, High-Fidelity, Frequency-Flexible Two-Qubit Fluxonium Gates with a Transmon Coupler, *Phys. Rev. X* **13**, 031035 (2023); arXiv:2304.06087v1 [quant-ph] 2023.
[39] Y. Guo, Q. Liu, W. Huang, Y. Li, T. Tian, N. Wu, S. Zhang, T. Li, Z. Wang, N. Deng, *et al.*, 29.4 a cryo-CMOS quantum computing unit interface chipset in 28nm bulk CMOS with phase-detection based readout and phase-shifter based pulse generation, in 2024 IEEE International Solid-State Circuits Conference (ISSCC), vol. 67, pp. 476–478, IEEE, 2024.
[40] A. Bazammul *et al.*, High-speed DAC/ADC Implementation Using RFSoC FPGA for Quantum Computing, Proceedings Volume 13581. Photonic Computing: From Materials and Devices to Systems and Applications II; 135810A (2025).
[41] K. H. Park *et al.*, ICARUS-Q: A Scalable RFSoC-Based Control System for Superconducting Quantum Computers, *Rev. Sci. Inst.* **93**, 104704 (2022).
[42] M. O. Tholén *et al.*, Measurement and control of a superconducting quantum processor with a fully integrated radio-frequency system on a chip, *Rev. Sci. Instrum.* **93**, 104711 (2022).
[43] K. Kang, B. Kim, G. Choi, S.-K. Lee, J. Choi, J. Lee, S. Kang, M. Lee, H.-J. Song, Y. Chong, *et al.*, A 5.5 mw/channel 2-to-7 GHz frequency synthesizable qubit-controlling cryogenic pulse modulator for scalable quantum computers, in 2021 Symposium on VLSI Circuits, pp. 1–2, IEEE, 2021.
[44] S. Huang *et al.*, Microwave Package Design for Superconducting Quantum Processors, *PRX Quantum* **2**, 020306 (2021).
[45] "Amumetal 4K" and "Cryoperm" Magnetic Shielding Materials, Data Sheet Number 20220503160050 (Amuneal Corp., 2011).
[46] T. Cai *et al.*, Multiplexed double-transmon coupler scheme in scalable superconducting quantum processor, arXiv:2511.02249v1 [quant-ph] (2025).
[47] L. C. Herrmann, Simulations and measurements of high-frequency transmons for a quantum transduction experiment, Master's Thesis, Swiss Federal Institute of Technology, 2024.
[48] V. Sivak *et al.*, Reinforcement learning control of quantum error correction, arXiv:2511.08493 [quant-ph].
[49] H. P. Nautrup *et al.*, Optimizing Quantum Error Correction Codes with Reinforcement Learning, *Quantum* **3**, 215 (2019).
[50] E. Hyyppä *et al.*, Reducing Leakage of Single-Qubit Gates for Superconducting Quantum Processors Using Analytical Control Pulse Envelopes, *PRX Quantum* **5**, 030353(2024).
[51] Guo, D., Yang, D., Zhang, H. *et al.* DeepSeek-R1 incentivizes reasoning in LLMs through reinforcement learning. *Nature* 645, 633 (2025).
[52] G. P. B. Huber *et al.*, Parametric multi-element coupling architecture for coherent and dissipative control of superconducting qubits, *PRX Quantum* **6**, 030313 (2025). [arXiv preprint arXiv:2403.02203, 2024].
[53] T. C. White *et al.*, Travelling Wave Parametric Amplifier with Josephson Junctions using Minimal Resonator Phase Matching, arXiv:1503.04364v1 [cond-mat.supr-con].
[54] R. Moretti *et al.*, Transmon qubit modeling and characterization for Dark Matter search, arXiv: 2409.05988v3 [quant-ph] 31 December 2024.

[55] J. Kang, C. Kim, Y. Kim and Y. Kwon, New design of three-qubit system with three transmons and a single fixed-frequency resonator coupler, *Sci. Rep.* **15**, 12134 92025).

**Supplementary Material:**

**Appendix A**

A simple, low-frequency (4-6GHz), and flux-tunable four transmon design is illustrated in the Figure SM1, an earlier experiment conceived by the author, presented here for completeness.

The parameters for the qubit transmons and resonators are designed in the proprietary EDA design software "AMR Microwave" (Cadence AMR Microwave Office, Cadence Corp.)

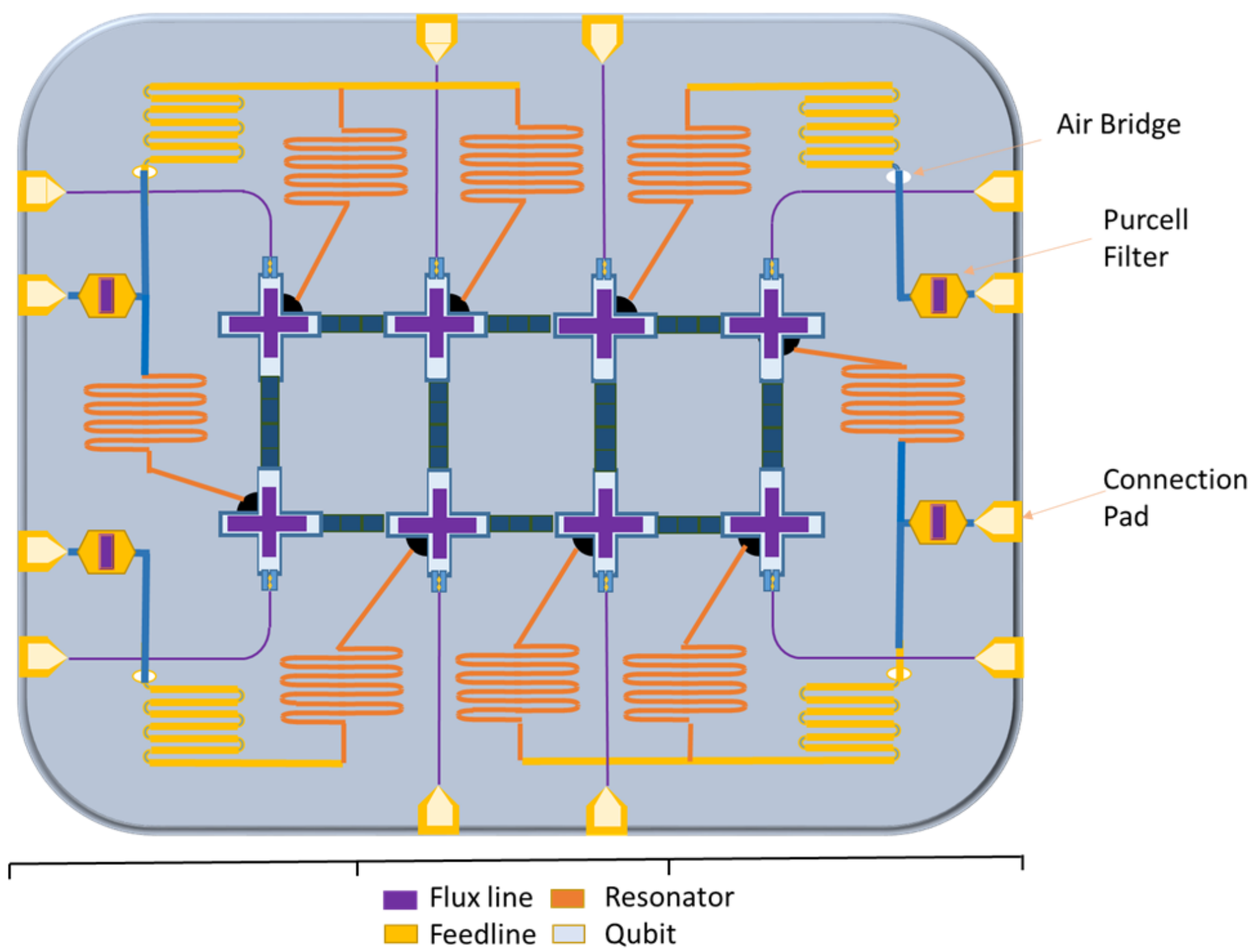

Figure SM1: A Schematic of a Low-Frequency Flux-Tunable Eight-Qubit Design, an Earlier Experiment.

.